\documentclass[comsoc,journal]{IEEEtran} 

\usepackage{cite,graphicx,amsmath,amssymb}
\usepackage{comment}
\usepackage{graphicx}
\usepackage{amsfonts,amssymb}
\usepackage{subfigure}
\usepackage{xcolor}
\usepackage{multirow}
\usepackage{url}
\usepackage{enumerate}
\usepackage{enumitem}
\usepackage[
top    = 1.70cm,
bottom = 0.82 in,
left   = 0.63 in,
right  = 0.63 in]{geometry}
\usepackage{xspace}
\usepackage{diagbox}
\newcommand\ahmad[1]{\textcolor{black}{#1}}
\newcommand\jiadong[1]{\textcolor{black}{#1}}
\newcommand\jiadongmr[1]{\textcolor{black}{#1}}
\newcommand{\one}{({\em i}\/)\xspace}
\newcommand{\two}{({\em ii}\/)\xspace}
\newcommand{\three}{({\em iii}\/)\xspace}
\newcommand{\four}{({\em iv}\/)\xspace}

\title{Bi-directional Digital Twin and Edge Computing in the Metaverse}
\author{Jiadong~Yu $^*$, Ahmad~Alhilal $^*$, Pan~Hui, Danny~H.K.~Tsang

\thanks{\textbf{ *~J. Yu and A. Alhilal contributed equally to this work~}
}
\thanks{J. Yu, P. Hui, and D.H.K. Tsang are with the Hong Kong University of Science and Technology (Guangzhou).}
\thanks{A. Alhilal, P. Hui, and D.H.K. Tsang are with the Hong Kong University of Science and Technology.}}
\begin{document}
\maketitle

\begin{abstract}
The Metaverse has emerged to extend our lifestyle beyond physical limitations. As essential components in the Metaverse, digital twins (DTs) are the real-time digital replicas of physical items. Multi-access edge computing (MEC) provides responsive services to the end users, ensuring an immersive and interactive Metaverse experience. While the digital representation (DT) of physical objects, end users, and edge computing systems is crucial in the Metaverse,  the construction of these DTs and the interplay between them have not been well-investigated. In this paper, we discuss the bidirectional reliance between the DT and the MEC system and investigate the creation of DTs of objects and users on the MEC servers and DT-assisted edge computing (DTEC). To ensure seamless handover among MEC servers and to avoid intermittent Metaverse services, we also explore the interaction between local DTECs on local MEC servers and the global DTEC on the cloud server due to the dynamic nature of network states (e.g., channel state and users' mobility). \jiadong{We investigate a continual learning framework for resource allocation strategy in local DTEC through a case study. Our strategy mitigates the desynchronization between physical-digital twins, ensures higher learning outcomes, and provides a satisfactory Metaverse experience.}
\end{abstract}

\section{Introduction}
The Metaverse is an extended reality (XR) that blends the physical world with the virtual world. In recent years, leading technology companies such as Facebook and Microsoft have highlighted the convenience and importance of novel scenarios in the Metaverse. 
As one of the most important components in the Metaverse, the digital twin (DT) is the digital replica that covers the life cycle of their physical counterparts, i.e., physical twin (PT), such as a physical object, process, or system~\cite{ladj2021knowledge}. With the explosive deployment of the internet of things (IoT) and the maturity of communication technologies, physical data can be collected in real-time to create DTs in the virtual world.
DT creation requires multi-access edge computing (MEC) that is physically closer to the data source (PT) to further reduce the unnecessary cloud traffic and ensures low latency in processing the data and maintaining the DTs. 
\jiadong{The International Organization for Standardization (ISO) formed a working group named ISO/IEC JTC 1/SC 41/WG6 to focus on DT standardization. The group is proposing a draft for a potential standard called ISO/IEC AWI 30173 Digital twin – Concepts and terminology\cite{iso_digitaltwin}. The core concept of the DT, however, has not reached a consensus and will be further clarified.} 

The implementation of DT helps in evaluating decisions without the need for physical testing\cite{corrado2022combining}. Likewise, the MEC ecosystem can be managed and troubleshot to deliver 
high-quality services to the end users using DT-assisted edge computing (DTEC). This results in a bidirectional reliance between DT and the MEC system. In this paper, we provide an overview of the digital components in the Metaverse.
After discussing the creation of DT and DTEC, we highlight some potential challenges and provide research directions to address them. This research pledges to answer the following two research questions (RQs):



\textbf{\jiadong{RQ1: How to create DTs of objects, users, and DTEC on MEC servers?}}

\jiadong{IoT sensors collect data from physical objects and users and send it to MEC servers to create their DTs in the virtual world. The DTEC receives the MEC system's up-to-date data (e.g., resource consumption, load, and performance). This transmission is accomplished via intra-twin communications. DTEC also retrieves the users' data from hosted DTs via inter-twin communications. The MEC servers provide the necessary resources (computing, cashing, and storage) to enable the DT construction which includes training its machine learning (ML) models. The decisions made by the DTs' models are returned as feedback to the physical counterparts (PTs) through intra-twin communications.}

\textbf{RQ2: How to ensure high and fair QoE in the Metaverse through the interplay between DTs, local DTECs, and global DTEC?}

The Metaverse users can display fields of view (FOV) in virtual space by wearing lightweight devices such as head-mounted displays (HMDs). The MEC servers are capable of rendering the requested FOVs and streaming them to the users. 
The user's DT, hosted on the MEC server, is updated as the user's eye gaze and head rotation change. This DT also obtains other performance information (e.g., latency, visual quality) which defines the user's quality of experience (QoE). The local DTEC can then retrieve the performance information from the user's DT, and the state of communication and computing resources to make decisions. These decisions assist the MEC system in automatically adjusting the system settings to ensure high QoE. However, the dynamic edge network states and network topology impact the provision of continuous Metaverse services. As such, a global DTEC is constructed on the cloud server to select the target MEC server for hosting the user's DT. This global DTEC retrieves information from local DTEC. It is also supported by the  entire edge network topology and an ML model to make DT migration decisions. 

\section{Digital Components and DT Applications in the Metaverse}
In this section, we first define the digital components of the virtual worlds and then introduce three main application cases of the DT in the Metaverse.
\subsection{Digital Components of the Virtual Worlds}
The Metaverse blends the physical world with the virtual world. As essential components in the virtual world, DTs are large-scale and highly accurate digital replicas of physical world objects and entities (avatars).  
The real-time data flow from PTs to DTs and the feedback from DTs to PTs tie the physical space with digital space.
The core thing of the meta-universe is not only the DTs, but the digital “natives”. Many things native to the digital world have no correspondence with the real world. 
\textit{Digital Natives} are digital creations generated by content creators and exist solely in the virtual world~\cite{DBLP:journals/corr/abs-2110-05352}.
They can be supported by interconnected ecosystems, including culture, economics, laws and regulations, and social norms. These ecologies facilitate the creation of concrete commodities and intangible contents and are comparable to the current rules and laws in the physical world society. Therefore, the DTs together with the digital natives make the Metaverse an independent, self-sustaining, and enduring virtual environment that can interact and coexist with the physical world. 

The difference between DT, digital natives, digital model, and digital shadow
resides in the automation of data flow (manual or automatic) between these digital components and their correspondents in the physical world. The \textit{Digital Model} is defined as the digital representation of an existing or anticipated physical item with manual data exchange with the physical world.  As a result, after the digital model has been produced, modifications made to the physical world have no automatic influence on the digital model. The \textit{Digital Shadow} refers to an automatic one-way flow from the physical world to the digital representation of an object, while the return data flow is manual. Conversely, the data flow, DT-to-PT, is automatic in both directions between the physical world and the digital world~\cite{ladj2021knowledge}.

\subsection{Real-world Applications}

\begin{figure*}[t]
  \centering
  \includegraphics[width=0.73\textwidth]{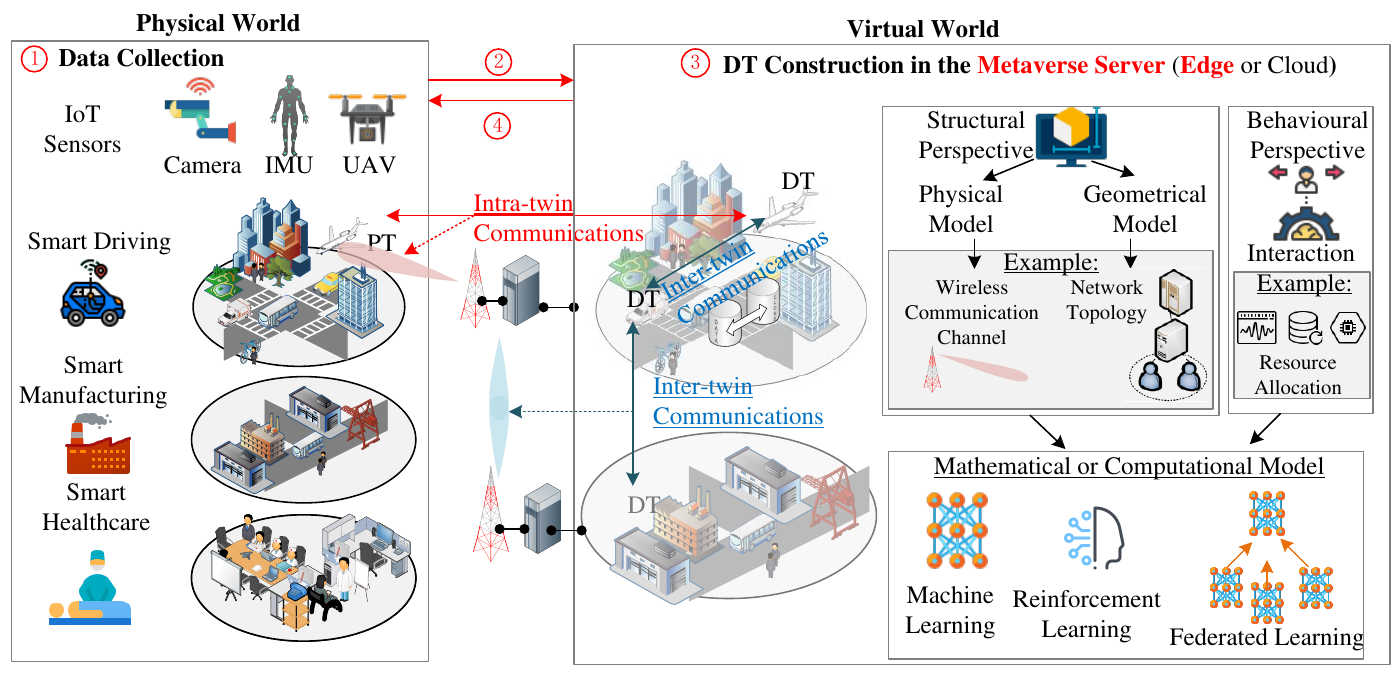}
  \caption{\jiadong{Edge computing and communication for DT creation in the Metaverse.}}
\vspace{-15pt}
  \label{DTpipeline}
\end{figure*}

As illustrated in Fig.~\ref{DTpipeline}, we briefly introduce three main application cases of the DT in the Metaverse with the support of IoT, i.e., smart driving, smart manufacturing, and smart healthcare.
For smart driving, traditionally, drivers are fully responsible for avoiding safety risks. With the use of the internet-of-vehicle (IoV), the vehicle's embedded sensors, and DTs, accidents can be predicted and augmented to the vehicle's display (e.g., on the windshield) with contextual information. For instance, it can show virtual objects representing the unsafely driven with position and speed information. This keeps the driver alert of any potential safety threats. Nissan company has put efforts to create a Metaverse platform called invisible-to-visible (I2V) with the augmented reality (AR) interface. 
For smart manufacturing, due to the maturity of technologies such as the Industrial Internet of Things (IIoT) and the emergence of DTs. There are four prominent use cases for applying DTs in the production line, i.e., product design, virtual workshop or factory, usage monitoring, and MRO (Maintenance, Repair, and Overhaul)~\cite{8258937}. 
For smart healthcare, as one of the main applications, remote healthcare is intended to offer patients timely, rapid, and accurate operations in emergency conditions and travel difficulties. Therefore, the timely and high-precision interaction between physical and virtual spaces is a necessity. DTs can enable distant operations~\cite{8632888}. For instance, Da Vinci system is a current cutting-edge robotic surgery technology to do prostate cancer surgery on patients using miniature robots. Doctors control the robot using a specialized controller interface while viewing the patient in VR. Additionally, IoT sensors and monitoring devices (attached to the patient's body) can feed the patient's DT. This DT allows doctors to assess the patient's health state. This includes patient lifestyle, medication usage, and emotional changes over time with severity forecast as DT behavior to enable early treatment.

\section{Edge Computing-Enabled DT Creation}
\label{section3}
\jiadong{Building a virtual representation of large-scale physical space requires extending the fundamental DT concept towards massive twinning~\cite{hashash2022edge}. 
Handling the intensive computation to maintain massive twinning (e.g., model training) requires cooperative edge-cloud computing. Edge servers provide real-time services where data processing must be performed in real-time, while cloud servers perform computationally heavy tasks and handle the integration of heterogeneous DTs between edge servers.} In this section, we study the DT creation pipeline to maintain DTs on the edge computing platform. 
\subsection{Real-time Data Collection from PTs} 
Fig.~\ref{DTpipeline} illustrates data collection from multiple sources (e.g., camera, inertial measurement units (IMU) suit). Data collection from the physical space is \textbf{the first step} to synchronize DTs with their PTs. This requires the collection and ingestion of a large volume of sensor data into the Metaverse computing platform (edge or cloud). IoT refers to the collective network of IoT devices (equipped with sensors) and the technology that facilitates communication with other objects. IoT devices have minimal quantities of computing and storage abilities.
With the wide spread of IoT, IoT devices can constantly sense and transmit sensing or pre-processed data to edge or cloud servers. These data can be fused and integrated and fed into the Metaverse server to ensure well-constructed and synchronized DTs in the virtual space. This enables plenty of functions such as monitoring, remote access, asset tracking, prediction, and interaction. Apart from conventional sensing systems, unmanned aerial vehicles (UAVs) can be utilized as movable IoT to sense and gather information about the states of the physical counterparts~\cite{9865226}.
\subsection{Communications for DT}
Fig.~\ref{DTpipeline} also depicts two types of communications, intra-twin communications (or so-called virtual-to-physical--V2P) and inter-twin communications (or so-called virtual-to-virtual--V2V)~\cite{9429703}. 
Intra-twin communication refers to the interaction between the DT and its physical counterpart PT. 
DTs and PTs exchange data and information through existing communication network technologies such as fiber links, WiFi, and 5G. Additionally, emerging sixth-generation (6G) technologies, heterogeneous radio frequencies such as sub-6 GHz, millimeter Wave (mmWave) and Terahertz (THz), intelligent reflecting surfaces (IRS), and non-orthogonal multiple access (NOMA) can provide high-capacity communications~\cite{yu20226g}.
PTs might move in the physical environment, and intra-twin communication, thus, must ensure heterogeneous wireless interfaces to ensure seamless communication. Fig.~\ref{DTpipeline} illustrates \textbf{the second step} to send data from PTs to servers of DTs in edge computing. Inter-twin communication describes the interaction among DTs which requires communication among their hosting servers and between DTs on the same server. Such communications can be realized through shared memory when communicating DTs are hosted on the same server or through communication networks between different servers. 
\subsection{DT Construction in the Virtual World}
\label{section3c}
\textbf{The third step} is to create high-fidelity virtual models to reproduce the geometry, physical properties, behaviors, and rules of the physical objects (PTs). These models (DTs) are maintained as components in the virtual world on the Metaverse servers.
There are mainly two perspectives, structural and behavioral perspectives, for the DT modeling~\cite{segovia2022design}.

From the structural standpoint, the DT is supposed to reflect the connection and assembly relations among the structures of the physical object or process. This includes simulating the physical and geometric models of the physical object or process. The physical model reflects the physical properties (e.g., speed, force, temperature, electromagnetic wave transmission), while the geometric model reflects the geometry properties such as shapes, colors, sizes, and relative positions of the items, the interfaces of the real system, and network topology.

From the behavioral standpoint, the DT is supposed to recreate the behavior of the physical process that the PT controls. This necessitates the use of a mathematical or computational model to connect variables of interest. For example, the model recreates the interaction between forces, acceleration, jerk, angular displacement, angular acceleration, and other phenomena in the physical process. Since understanding physical processes might not be feasible in complex systems, mathematical models are not possible. Data-driven behavioral models are more flexible and suitable, especially with the widespread of IoTs.

\subsection{DTs Feedback and PTs Update}

\textbf{The fourth step} is to provide feedback from DT to its PT. With DT's structural and behavioral awareness, they can inform decisions to its physical counterpart which requires establishing a feedback channel, intra-twin DT-to-PT communications.  
One example of behavioral DT for smart driving is to forecast the collision risk and feedback to the drivers as warning and safety instructions
. Accordingly, drivers can react on the roads to avoid potential dangers while driving. For smart manufacturing, the DTs can constantly send feedback as information to, for instance, forecast product testing, evaluate the different manufacturing strategies, predict the product's remaining life, and provide fault diagnosis~\cite{8258937}. The manufacturer can react accordingly, for instance, by improving the product design, adjusting the manufacturing process, and keeping an eye on the monitoring logs. For smart healthcare, the feedback can be conveyed as visual information aligned with haptic forces to the doctors, thus enabling remote surgery operation~\cite{8632888}.

\section{DT-assisted Edge Computing}
\label{sec:edgecompute}
In this section, we first discuss the bi-directional reliance between the DT and the MEC system, and the three hierarchical DTs in the Metaverse. We then study the DTEC creation pipeline and investigate the behavioral models of local and global DTECs. Lastly, we provide a case study of local DTEC supported with a proposed continual learning framework.

\begin{figure*}[t]
  \centering
  \includegraphics[width=0.73\textwidth]{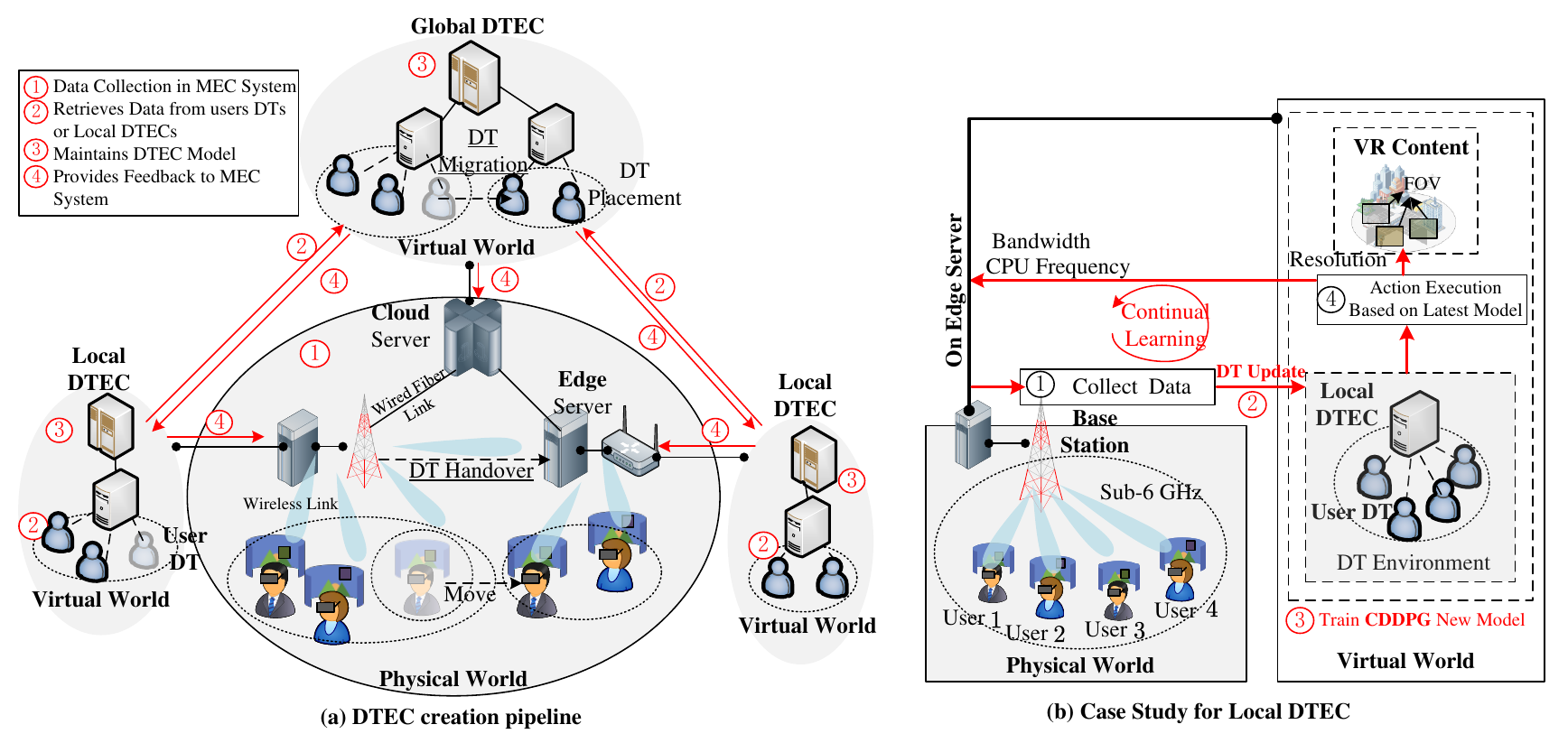}\\
  \caption{\jiadong{DTEC creation pipeline, the interplay between local DTEC and global DTEC, and a case study for local DTEC.}} 
    \vspace{-15pt}
  \label{DTEC}
\end{figure*}

\subsection{Why DTEC for the Metaverse?}
\label{section4a}

After creating and deploying massive twinning on Metaverse servers (edge or cloud), the Metaverse system must allow the users to access it through lightweight devices (e.g., HMDs, AR glasses, gloves) while providing an immersive user experience. However, these devices have limited resources (e.g., battery and computing power). As such, resource-intensive tasks such as FOV prediction, image processing, and rendering cannot be directly executed locally.
Edge computing allows Metaverse users to share communication resources (i.e., spectrum) and edge computing resources (i.e., caching memory and CPU frequency).

DTEC is the DT representation of the edge computing system. It can monitor the most up-to-date state of edge computing, real-time resource utilization, and network performance. DTEC can also fabricate emulation of users, load, and resource allocation to produce emulated data. These data help to train the DT model to find the optimal strategy for resource allocation. Accordingly, DTEC can make decisions and send feedback to the edge computing system to assign the resources to the users that ensure fairness QoE. The resources 
can be reserved for certain model training and Metaverse applications or users in advance. Moreover, DTEC can also continuously monitor the condition of the whole system, detect the unfairness of the resource distribution, or forecast any system failure in edge computing, thus enabling proactive intervention. 
We thus introduce three hierarchical DTs, i.e., User' DT, local DTEC, and global DTEC, that assist in immersive and fairness QoE in the Metaverse.
\begin{itemize}[leftmargin=1em]
\item \textbf{User's DT}: 
As the user's DT is placed at the edge server, the user transmits user tracking information  (e.g., head rotation and eye gaze) to predict the FOV which contributes to the behavior of the user's DT. The user also reports the QoE performance to the corresponding DT (via intra-twin communication).
\item \textbf{Local DTEC}: 
The local DTEC is placed on the MEC server and retrieves the information from the user's DT via inter-twin communications. The DTEC models the communication and computation state of the local MEC system (e.g., CPU cycles and the frequency band and bandwidth) that are occupied for the user's FOV prediction and rendering. These models are constantly updated and calibrated (via intra-twin communications between the local DTEC and the MEC system).

\item \textbf{Global DTEC}:
The global DTEC is placed on the cloud server to maintain the topology of the entire edge network. This DTEC retrieves all the information of local DTECs via inter-twin communications. The global DTEC models the server selection strategy. This strategy defines the MEC server to host and serves the end user. It decides the target server based on the local DTEC's (e.g., latency, and load) which may result in DT migration to ensure high QoE.    

\end{itemize}

\subsection{DTEC Creation}
\label{section4b}
Fig.~\ref{DTEC}(a) illustrates the DTEC creation pipeline with two MEC servers interconnected with a cloud server. 
Following the DT construction pipeline introduced in Section~\ref{section3}, the construction of DTEC also includes four steps.
Firstly, the local DTEC collects real-time data on MEC states (e.g., resource utilization, enabled spectrum), which involves intra-twin communications. Secondly, it retrieves network and performance data of mobile end users (e.g., network throughput, visual quality, motion-to-photon latency) using hosted users' DTs, which involves inter-twin communications.
In the third step, it constructs the structural model (see example in Fig.~\ref{DTpipeline}, wireless communication channel, and network topology) and behavioral model that is trained through ML algorithms, federated learning (FL)~\cite{ramu2022federated}, or reinforcement learning (RL) to decide the optimal resource allocation. 
Lastly, the local DTEC returns the decision as feedback to the MEC system (via intra-twin communications). For instance, it adjusts the spectrum and bandwidth, increases or decreases the computing capacity assigned to a user, or redistributes the caching quota. However, massive twinning and cooperative edge-cloud computing requires a global DTEC. The global DTEC is maintained on the cloud server. The main difference between constructing local DTEC and global DTEC resides in the second step, in which the global DTEC retrieves the state of the MEC system from local DTECs via inter-twin communications. It then constructs the edge network as an entire structural model (i.e., global network topology). Note that both local network topology and global network topology change upon handover (transferring data connectivity from one edge server to another). The handover leads to a change of user DT's hosting server (later subsection discusses DT placement and DT migration). 

\subsection{DTEC Behavioural Perspective}
In this following subsection, we discuss two behavioral aspects of DTEC construction (i.e., DT placement and migration in global DTEC, and resource allocation in local DTEC).

\subsubsection{DT Placement and DT Migration in global DTEC}
The mobility of end users impacts the provision of continuous MEC services when using a single local DTEC, affecting the real-time experience of user interaction in the Metaverse. This phenomenon is exacerbated with the dynamic network topology (e.g., handover across edge servers) and dynamic MEC network states (e.g., available computation resources and channel states). Therefore, it is crucial to consider the system's latency and energy consumption of DT maintenance in global DTEC.
Lu et al.~\cite{9491087} formulate an edge association problem to reduce the average system latency and improve user utility in global DTEC. The problem is divided into DT placement and DT migration, based on the various running stages. As illustrated in Fig.~\ref{DTEC}(a), DT placement in global DTEC (located at the cloud server) finds the optimal server association that is suitable to users' current condition. However, the mobility of mobile users may lead to handover across MEC servers. As reconstructing the user's DT on the target server is time and resource-consuming, migrating the user's DT from the source hosting server to the target (i.e., DT migration) is considered a more efficient solution.

\subsubsection{Resource Allocation in local DTEC} 
As discussed in Section~\ref{section4a}, DT construction and immersive Metaverse experience require the support of the MEC system. To ensure the MEC system operates in an efficient resource-utilization way for both cases, it is necessary to rely on local DTEC which can assist in evaluating the performance of the edge computing network and making decisions for customized and fair resource allocation (e.g., communication and computing resources). 

From the DT construction perspective, when leveraging FL to construct DT models for PTs based on the data in the physical world, local training at the PTs and model aggregation at the edge serve can still be time and energy-consuming. 
Thus, it is necessary to optimally allocate the communication (i.e., transmission power) and computation resources (e.g., CPU cycles at both PTs and edge servers) in the MEC system to improve efficiency (utilization of resources) during the training stage ~\cite{9145588}.

From the immersive Metaverse experience, computation-intensive tasks (e.g., FOV prediction and high-quality virtual scene rendering) are offloaded to edge servers for fast computing and return rendered video streams to the users' devices for display. Based on the QoE performance of the users, the local DTEC can provide feedback to the MEC system to automatically adjust the computing capacity assigned to a user and redistribute the caching quota. Local DTEC can also provide feedback to the local MEC system to increase or decrease the bandwidth assigned to a specific user or activate another frequency band to upgrade the channel upon congestion. 

\subsection{\jiadong{Case Study: Local DTEC for high QoE in the Metaverse}}


\ahmad{Conventional approaches have utilized RL algorithms to advise the resource allocation in MEC systems. The RL agent interacts with the physical environment, obtains observations that define its current state, and rewards or penalizes its action in the previous episode. This association of states, actions, and rewards builds the RL policy that guides its decision-making to achieve the highest expected cumulative reward over time. However, it is noteworthy that this constant interaction introduces considerable communication overhead that can cause high latency, leading to desynchronization between the real and digital worlds. Simulated systems attempt to mimic these interactions, but they are unable to fully reproduce the complexities of the actual world, particularly in dynamic contexts. These simulations overlook the dynamic nature of the user's environment, which has a detrimental effect on the resource allocation strategy.}

\jiadongmr{To cope with this, we propose a continual RL framework that continuously and directly interacts with the DT to learn the optimal up-to-date strategy for resource allocation. Fig.~\ref{DTEC} (b) emphasizes three aspects to consider in the framework:}
\begin{itemize}[leftmargin=5mm]
    \item \jiadongmr{Physical World: We conduct experiments on $4$ VR users accessing VR content on a local MEC server. The edge server maintains its DT (local DTEC) to allocate its CPU resources ($15\text{GHz}$), and bandwidth resources ($10\text{Mbps}$) while maximizing the QoE of the whole system\footnote{\jiadongmr{To be noticed, the setting of $15\text{GHz}$ CPU resources are the combination of multiple computation resources. These resources can work together to execute multiple tasks in parallel, improving overall CPU frequencies. More information about the system and its configurations can be found in~\cite{yu2023attention}}}.  }
    \item \jiadongmr{Virtual World: Following the hierarchical DTs mentioned in Section~\ref{section4a}, there are users' DT and local DTEC. The users' DT records both the real-time attention-based eye-tracking data and the individual QoE performance. Rendering and downloading latencies, and resolution of rendered VR content are considered the primary pillars of QoE. The local DTEC reflects the transmission throughput and CPU frequency for each user and time-horizon fairness QoE among users. The behavioural DTEC makes resource allocation decision(i.e., adapting the attention-based resolution of VR content, allocating CPU resources, and allocating bandwidth resources) (Step 3).}
    \item \jiadongmr{Intra-twin Communications: There are mainly two categories of interaction, one for updating the DT from the physical world to the virtual world (Step 2) and another for broadcasting the behavioral decision from the virtual world to the physical world (Step 4). With the support of this intra-twin communication, the DT environment is constantly updated and the physical world can execute the up-to-date strategy that mitigates the desynchronization problem. }
\end{itemize}

\ahmad{In our continual RL framework, the local DTEC constantly updates its hierarchical DTs (step 2) and continues to interact with the real-time DT environment (step 3) to learn the decision-making behavior. We use deep deterministic policy gradient (DDPG) as the core RL algorithm. It is renowned for its capacity to handle continuous action spaces and is employed here to guide the agent's decision-making. In contrast to the pre-trained DDPG-based frameworks, the proposed framework, i.e., continual DDPG (CDDPG), can support local behavioral DTECs in making timely decisions for optimal resource allocation.}

\begin{figure}[t]
  \centering
  \includegraphics[width=.35\textwidth]{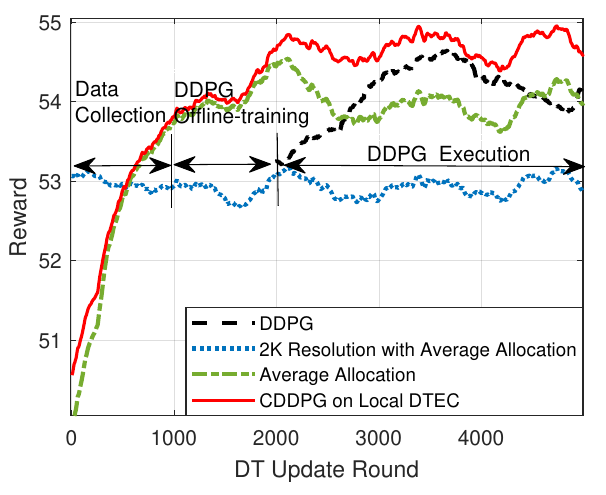}
  \caption{\jiadong{Reward convergence per DT updates. CDDPG exhibits the highest performance, compared to the benchmarks.}}
      \vspace{-15pt}
  \label{figa}
\end{figure}

\jiadong{We evaluate our framework against multiple benchmarks: 1) 2K Resolution with Average Allocation which assigns 2K resolution for different attention levels, and CPU frequency and bandwidth uniformly to the $4$ users; 2) Average Allocation which assigns the uniformly CPU and bandwidth resources to the $4$ users under adaptive attention-based resolution; and 3) Conventional, non-continual learning, RL-based method (DDPG). Specifically, the collected data during the first $1000$ rounds of DT update is used for DDPG offline training. The actual allocation starts at the $2000$ rounds. Fig.~\ref{figa} illustrates the convergence performance of our framework compared to the benchmarks over time. The DDPG execution outperforms both the 2K Resolution with Average Allocation and Average Allocation most of the time after $2000$ DT update rounds. However, our proposed CDDPG on local DTEC shows superior performance compared to all benchmarks. Being evolvable using new DT data, the continual learning framework is more effective.}

\begin{figure}[t]
  \centering
  \includegraphics[width=.35\textwidth]{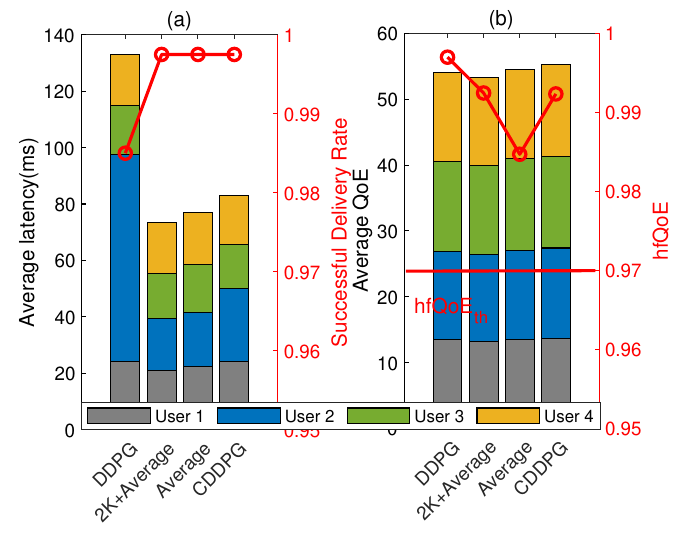}
  \caption{\jiadong{(a) Average latency, successful delivery rate. (b) average QoE and horizon fairness QoE for 4 users. CDDPG achieves the best trade-off between low latency and high delivery rate, and the highest QoE, while ensuring fairness QoE, compared to the benchmarks. }}
    \vspace{-15pt}
  \label{figb}
\end{figure}

\jiadong{Fig.~\ref{figb} presents the performance of the proposed framework.
Fig.~\ref{figb} (a) illustrates the average time latency (\jiadongmr{i.e., Rendering and downloading latencies}) and the successful delivery rate of the algorithms (\jiadongmr{i.e., the time latency is within the $150\text{ms}$ time threshold}). DDPG which disregards the recent DT update exhibits the highest time latency and lowest successful delivered rate ($98.5\%$), while the algorithms with DT update have a higher successful delivery rate ($99.75\%$). Among all the benchmarks, 2K Resolution with Average Resource Allocation shows the lowest time latency, which can be attributed to the smaller data size of the group of pictures (GOP) for rendering. Fig.~\ref{figb} (b) compares the average QoE and horizon fair QoE (hfQoE) of the algorithms. According to our reward function, all benchmarks satisfy the horizon fairness QoE threshold $\text{hfQoE}_{th} (0.97)$. Our proposed CDDPG achieves the highest QoE while meeting the fairness requirements. }

\ahmad{The focal point of this case study revolves around the enhancement of QoE concerning real-time VR content delivery. However, the conceptualized framework for continuous learning, rooted in a DT system, could be applied in various scenarios demanding instantaneous decision-making. Scenarios where engaging with the physical environment incur substantial costs such as communication and power consumption costs. For example, Fig.~\ref{DTpipeline} illustrates examples of smart driving, smart manufacturing, and smart healthcare. For smart driving system, with autonomous vehicles that requires real-time decision-making to navigate safely and efficiently. The proposed framework could be applied to enhance QoE in managing traffic flow, coordinating vehicle movements, and preventing accidents, particularly in urban environments with complex interactions. For smart manufacturing, real-time decisions are crucial for optimizing production efficiency. The proposed framework could help in situations where human intervention is costly due to the need for physical presence or where communication delays affect operations. For smart healthcare, real-time decision-making has a significant impact on the quality of patient care and the outcomes of medical treatments. This framework can be adapted to enhance the experience of diagnosing and treating patients remotely, especially in situations where physical access to medical facilities is limited or medical treatment is urgent and emergency.}
\section{Challenges and Research Directions}
The creation of DTs and DTEC will face multiple challenges related to the synchronization gap between twins, DT migration, and catastrophic forgetting in DTs.
\subsection{Minimal Twins De-synchronization}
Capturing real-time data from human users, physical objects, processes, and systems helps to construct high-fidelity massive twinning. This would enable a reality-like Metaverse environment. However, due to the dynamic environment (e.g., variable signal strength, and dynamic network throughput and topology), the de-synchronization gap between the twins (PTs and DTs) is expected to increase dramatically.
Designing a dynamic ML training process would assist in ensuring a minimal de-synchronization gap. 
This would require a DNN with a dual objective optimization function that jointly minimizes the loss function and the corresponding de-synchronization time. Besides, reducing the transmission latency would be an additional aspect to reduce this gap. Deploying cutting-edge communication technologies and optimizing resource allocation using DTEC would contribute to minimizing this gap. 

\subsection{Seamless DT Migration}
The dynamic environment is expected to impose variable delay and extra power consumption upon DT migration when handover occurs from one MEC server to another. This issue is exacerbated in the presence of heterogeneous capabilities and connectivity of MEC servers (e.g., GPU, memory, and communication link).
Seamless migration of DTs is essential to the provision of continuous Metaverse services. While DT migration requires transferring the up-to-date DT state and the historic states, an evolutionary migration approach would be an interesting direction to investigate. This approach should instantly send the up-to-date state and DT model's parameters, and then historical data can be sent on a priority basis. This approach ensures the continuity of the service and avoids unnecessary power consumption. 


\subsection{\jiadong{Catastrophic Forgetting in DTs}}
\jiadong{Learning-based DTs are susceptible to catastrophic forgetting. 
This typically happens when the agent updates its function approximation based on new experiences, causing it to forget or overwrite previously learned knowledge. 
For instance, if a DT is trained for specific operating conditions and then deployed in a different environment, its performance may deviate from the expected because it lacks the ability to generalize its knowledge to new conditions.
Several techniques could be used to cope with this phenomenon such as transfer learning and continual learning. Transfer learning involves leveraging pre-trained models to accelerate the learning process, while 
continual learning would ensure to learn of the new knowledge while not forgetting the prior knowledge (e.g., latent parameter storage, distillation-based knowledge retention, and rehearsal-based knowledge retention), allowing the DT to perform optimally in the presence of time-varying conditions and environment.}

\section{Conclusion}
This paper investigated the creation of digital replicas of physical objects, users (DTs), and MEC systems (DTECs) in the Metaverse.
The pipeline of DT creation includes four steps: \one data collection from physical twins (PTs), \two communication between DTs and PTs (inter-twin communications), \three DT logic as a structural and behavioral model on the MEC server, and \four feedback from DTs to PTs (intra-twin communications). DTEC construction follows a similar pipeline with different behavioral models. 
\jiadong{This paper also studied the interaction among DTs, and local and global DTECs to ensure a high and fair QoE in the Metaverse by proposing three hierarchical DTs (i.e., user' DT, local DTEC, and global DTEC). We evaluated a continual learning framework in a local DTEC over a case study of VR content streaming. Our finding revealed that the framework achieves the highest QoE for users and preserves time horizon QoE fairness. Since the DTEC behavioral model considers the latest DT updates, the local DTEC can assign the resources optimally to the Metaverse users.
Finally, we raised some challenges that the DT construction would face and highlighted research directions accordingly, specifically, ensuring minimal twins de-synchronization, seamless DT migration, and mitigating catastrophic forgetting.}

\section{Acknowledgement}
This work is supported in part by the Guangzhou Municipal Science and Technology Project under Grant 2023A03J0011, Guangdong Provincial Key Laboratory of Integrated Communications, Sensing and Computation for Ubiquitous Internet of Things, National Foreign Expert Project, Project Number G2022030026L, and the MetaHKUST project from HKUST (Guangzhou).

\bibliographystyle{IEEEtran}
\bibliography{IEEEabrv,mybib}
\renewenvironment{IEEEbiography}[1]
  {\IEEEbiographynophoto{#1}}
  {\endIEEEbiographynophoto}
  \vskip -2.5\baselineskip plus -1fil
\begin{IEEEbiography}{Jiadong Yu} (jiadongyu@hkust-gz.edu.cn) is currently a Postdoc Fellow at the Internet of Things thrust at the Hong Kong University of Science and Technology (Guangzhou). 
\end{IEEEbiography}
\vskip -2.5\baselineskip plus -1fil
\begin{IEEEbiography}{Ahmad Alhilal} (aalhilal@connect.ust.hk) is currently a Postdoc Fellow at the Hong Kong University of Science and Technology, IPO, Division of EMIA.
\end{IEEEbiography}
\vskip -2.5\baselineskip plus -1fil
\begin{IEEEbiography}{Pan Hui}(panhui@ust.hk) is a Chair Professor of Computational Media and Arts thrust at the Hong Kong University of Science and Technology (Guangzhou), and a Chair Professor of Emerging Interdisciplinary Areas and Director of the HKUST-DT Systems and Media Lab, Hong Kong University of Science and Technology.
\end{IEEEbiography}
\vskip -2.5\baselineskip plus -1fil
\begin{IEEEbiography}{Danny Hin-Kwok Tsang}(eetsang@ust.hk) is the Thrust Head and Professor of the Internet of Things Thrust at the Hong Kong University of Science and Technology (Guangzhou), and an Emeritus Professor of the Department of Electronic and Computer Engineering at the Hong Kong University of Science and Technology.
\end{IEEEbiography}
\end{document}